\documentclass{vgtc}                          

\graphicspath{{figures/}{pictures/}{images/}{./}} 

\usepackage{times}                     

\usepackage{tabu}                      
\usepackage{booktabs}                  
\usepackage{lipsum}                    
\usepackage{mwe}                       

\usepackage{mathptmx}                  

\onlineid{0}

\vgtccategory{Research}

\vgtcinsertpkg

\title{"I Wanted to Create my Ideal Self": Exploring Avatar Perception of LGBTQ+ Users for Therapy in Virtual Reality
}

\author{Anish Kundu\thanks{e-mail: anish@kmd.keio.ac.jp}\\ %
        \parbox{1.4in}{\scriptsize \centering Keio University   Graduate School of Media Design} %
\and Giulia Barbareschi\thanks{e-mail: barbareschi@kmd.keio.ac.jp}\\ %
     \parbox{1.4in}{\scriptsize \centering Keio University   Graduate School of Media Design} %
\and Midori Kawaguchi\thanks{e-mail: midori@kmd.keio.ac.jp}\\ %
     \parbox{1.4in}{\scriptsize \centering Keio University   Graduate School of Media Design} %
\and Yuichiro Yano\thanks{e-mail: yano.yuichiro@jichi.ac.jp}\\ %
     \parbox{1.4in}{\scriptsize \centering Juntendo University Faculty of Medicine} %
\and Mizuki Ohashi\thanks{e-mail: mizukion@belle.shiga-med.ac.jp}\\ %
     \parbox{1.4in}{\scriptsize \centering Shiga University of Medical Science} %
\and Kaori Kitaoka\thanks{e-mail: kitaoka@belle.shiga-med.ac.jp}\\ %
     \parbox{1.4in}{\scriptsize \centering Shiga University of Medical Science} %
\and Aya Seike\thanks{e-mail: seike.aya.w77@kyoto-u.jp}\\ %
     \parbox{1.4in}{\scriptsize \centering Faculty of Sports and Health Science, Ritsumeikan University} %
\and Kouta Minamizawa\thanks{e-mail: kouta@kmd.keio.ac.jp}\\ %
     \parbox{1.4in}{\scriptsize \centering Keio University   Graduate School of Media Design}}

\teaser{
  \centering
  \includegraphics[width=\linewidth]{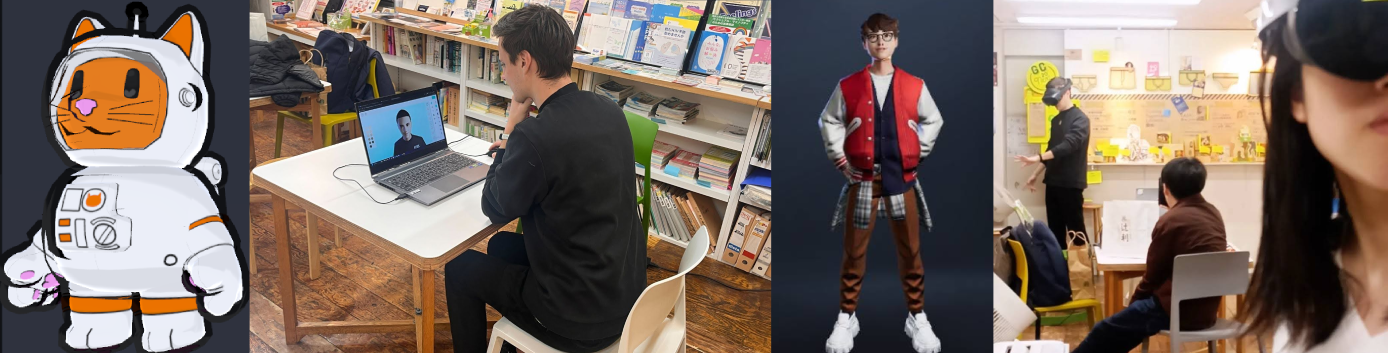}
  \caption{From left to right: Default avatar selected by a user, user creating a personal avatar, a created avatar, user during the experiment in VR space}
  \label{fig:teaser}
}

\abstract{
   In this paper we explore the potential of utilizing Virtual Reality (VR) as a therapeutic tool for supporting individuals in the LGBTQ+ community, who often face elevated risks of mental health issues. Specifically, we investigated the effectiveness of using pre-existing avatars compared to allowing individuals to create their own avatars through a website, and their experience in a VR space when using these avatars. We conducted a user study (n=10) measuring heart rate variability (HRV) and gathering subjective feedback through semi-structured interviews conducted in VR. Avatar creation was facilitated using an online platform, and conversations took place within a two-user VR space developed in a commercially available VR application. Our findings suggest that users significantly prefer creating their own avatars in the context of therapy sessions, and while there was no statistically significant difference, there was a consistent trend of enhanced physiological response when using self-made avatars in VR. This study provides initial empirical support for the importance of custom avatar creation in utilizing VR for therapy within the LGBTQ+ community.
} 


\begin{document}
\maketitle
\section{Introduction}

Members of the LGBTQ+ community experience discrimination and structural inequalities. Previous studies in HCI and beyond have shown that technology has the potential to support and empower the LGBTQ+ community by providing a system for mutual support, or giving the ability to create dedicated safe spaces, as well as help to build alliances between members of the LGBTQ+ community and others \cite{li_we_2023}.
In the above context, we raise the following research questions:
\begin{itemize}
    \item \textbf{RQ1:} How can an empirical examination provide insights on the difference between creating personalized avatars and selecting from a pre-determined list of avatars for VR Therapy?
    \item \textbf{RQ2:} What insights can be gleaned from analyzing physiological signals (specifically heart rate variability) during both avatar creation and utilization in VR in the above context?
\end{itemize}
To answer these questions, we conducted an experiment at an LGBTQ+ community center, where participants engaged with VR using pre-selected avatars and self-made avatars. Our findings show participants preferred to create their own avatars for therapy in VR. Our main contributions include: (i) A community-driven empirical examination of avatar creation and embodiment in VR; (ii) An analysis of heart rate variability in the above context.

\section{Background and Related Works}
\subsection{Social VR \& VR therapy in the LGBTQ+ community}
VR has been shown to have the potential to positively impact the LGBTQ+ community, especially in the context of gender identity, as it can help challenge stereotypes by the use of non cis-gendered avatars \cite{10.1145/3383652.3423876}.
Social VR can be described as any form of a multi-user VR space where people can interact and have social support, which is important for handling stressful situations \cite{cobb_social_1976}, particularly within the LGBTQ+ community. Social VR has demonstrated its ability to offer novel affordances to communities, including enhanced inclusivity \cite{10.1145/3432938}. As an example, previous work has highlited the opportunities it affords to explore diverse gender identities online \cite{freeman_rediscovering_2022}.
In this context, social VR platforms hold promise in assisting therapists to create inclusive virtual environments where LGBTQ+ individuals feel comfortable expressing themselves authentically, enhancing the therapeutic experience and promoting mental well-being. Though still in it's early stages, design considerations have been given to “queer visibility” in social VR \cite{freeman_acting_2022}. Moreover, within the broader context of Digital Therapy, which encompasses VR, exploration has identified six key themes: Simple delivery, Flexible use, Seamless Interactivity, Personalization, and Support \cite{bolesnikov_queering_2022}.

\subsection{Personalized avatars and embodiment in therapy and social VR}
Latoschik \cite{latoschik2017effect} and Wolf \cite{wolf2021embodiment} show that photorealistic avatars can enhance embodiment and body ownership but may also cause discomfort and affect self-image. Gall \cite{gall2021embodiment} found that increased embodiment heightens emotional responses, improving effectiveness but also vulnerability.
Freeman \cite{freeman2020my} emphasizes the importance of personalized avatars for self-representation and identity in social VR. Zhang \cite{10.1145/3491102.3517624} and Mack \cite{10.1145/3517428.3544829} highlight how this is particularly significant for marginalized groups, like those with disabilities, for asserting control over their narratives. These findings suggest that exploring avatar personalization with the LGBTQ+ community is a crucial research direction.
\section{Methodology}
The experiment was conducted in a Akta community center, a safe space for LGBTQ+ people available free of charge to anyone, located in Tokyo's LGBTQ+ district  'Shinjuku ni-chōme', which boasts the largest concentration of LGBTQ+ bars in the world. The author approached visitors and inquired if they would be interested in trying out VR and participate in the experiment, the inclusion criteria was simply being a part of the LGBTQ+ community. Visitors who couldn't attend but expressed interest were informed of a future experiment date, and researcher shared contact details for those wishing to join. The experiment spanned two days, with 3 participants on the 1st day and 7 on the 2nd.
\subsection{Experiment Structure}
The study was approved by the ethics committee of Shiga University of Medical Science (RRB23-047). Participants were provided a thorough explanation of the research process and informed that all data collected would be completely anonymized.
\begin{table}[ht]
\centering
\caption{Experimental procedure and times}
\small 
\begin{tabular}{|p{0.7\linewidth}|p{0.2\linewidth}|}
\hline
\textbf{Activity} & {Time} \\
\hline
Experiment Enrollment and VR familiarization & 10 minutes \\
VR discussion with pre-selected avatar & 15 minutes \\
Avatar creation & 20 minutes \\
VR discussion with custom-made avatar & 15 minutes \\
Questionnaire response & 10 minutes \\
\hline
\end{tabular}
\label{tab:flow}
\end{table}
\newline \textbf{Stage 1: Enrollment and VR familiarization}
\newline After a description of the experiment flow, participants were asked to wear a FitBit Inspire 3 device and asked to fill in a pre-survey for collecting Ethnographic data. The author then entered a VR space with the participants and taught them how to use the VR interface. Two Meta Quest Pro, loaded with a commercially available preinstalled application VRChat\footnote{\url{https://hello.vrchat.com/}} were used. The author made sure to use a private VR space accessible only to the participant and the researcher. The participants bodies were tracked though the HMD position and native hand tracking of the Oculus Quest Pro and participants could move around the virtual space using a hand gesture. The participant and researchers stood  physically apart from each other by around 5m in the centre so as to not break immersion. The virtual space was a large white room, around 5m by 3m, with one wall having a floor to ceiling mirror for the participants to see their avatar.
\newline \textbf{Stage 2: VR discussion with pre-selected Avatar}
\newline Participants chose an avatar from a default list of about 30 options inside VRChat, ranging from anthropomorphic to non-humanoid designs. After selection, a researcher in a generic humanoid avatar conducted a semi-structured interview, focusing on the reasons for avatar choice, perceived differences from their real body, and willingness to use the avatar for mental health therapy in VR.
\newline \textbf{Stage 3: Avatar Creation}
\newline Participants removed the headset and created their own avatar using Ready Player Me\footnote{\url{https://readyplayer.me/avatar}} on a Galleria XL7C-R36 laptop after a brief questionnaire. This free online tool generates a customizable full-body avatar from a photo, allowing changes to facial features, hairstyle, clothing, and accessories. Participants could use their own photo or one from the internet. After the initial avatar was generated, the researcher left the participant to customize it independently.
\newline \textbf{Stage 4: VR discussion with custom-made avatar}
After creating their avatar, participants took a 10-minute break for Fitbit data collection while the avatar was uploaded to VRChat. The researcher then conducted a second semi-structured interview in VR, focusing on differences from the previous session, reasons for customization choices, and comparisons between custom and pre-selected avatars. These conversations were also documented for later analysis.
\newline \textbf{Stage 5: Questionnaires}
\newline Participants were then asked to fill in a longer questionnaire about the entire experiment. They were asked to spend 10 minutes to respond to the questions and take a short break for another recording of the rest time physiological signals.
\newline \textbf{Participant Data}
In the experiment, we enrolled a total of 10 participants from diverse backgrounds: 7 from Japan, and 1 each from Vietnam, the United States, and Germany. 6 identified as male and 4 as female, with 3 individuals self-identifying as non-binary/gender-queer. 7 participants identified as homosexual, with 1 participant as asexual and 2 declined to disclose. Age-wise, 7 participants were aged 20-29, while 3 were aged 30-39.
\subsection{Data Collection and Analysis}
\subsubsection{Semi-structured Interviews}
We adopted the thematic analysis approach \cite{braun2006using} for qualitative analysis. After transcribing the VR conversations between participants and the researcher, the primary author organized the interview responses into a table for discussion with co-authors. By identifying prevalent themes, we discussed their significance and structured them into a cohesive narrative. The short 15-minute discussion period was chosen to ensure that participants' answers reflected their initial thoughts, as well as maintaining their comfort as visitors to the community center as a safe space.
\subsubsection{Physiological Data}
To monitor participants' physiological responses and measure the effect of VR experiences with both pre-existing and custom avatars on Heart Rate Variability (HRV), participants wore FitBit Inspire 3 devices throughout the experiment. These devices, which use photoplethysmogram technology at a 25Hz sampling frequency, have been validated for detecting affective states in various tasks \cite{yan2020affect}. Participants were assigned unique IDs, and the data was updated to the SelfBase platform \footnote{https://www.technology-doctor.com/selfbase} for analysis.

Data was partitioned into six segments corresponding to the activities in Table \ref{tab:flow}. For each segment, we extracted heart rate, CVRR, and SDNN, calculating descriptive statistics (mean, median, SD, min, max, Q1, Q3, and 95\% confidence intervals). These measures evaluate the balance between the sympathetic and parasympathetic systems, indicating arousal states. Low HRV is linked to sympathetic activation (e.g., fear, excitement), and high HRV to parasympathetic activation (rest, recovery). SDNN reflects total variability between heart beats over a period, with higher values indicating lower arousal \cite{mohammadpoor2023arousal}. CVRR, the ratio of SD to mean R-R intervals, is sensitive to autonomic activity changes, with higher values indicating lower arousal \cite{li2022heart}.

No missing data was recorded, and normality was confirmed using the Shapiro-Wilk test. A Repeated Measures ANOVA was performed to identify  variations between segments (significance level 0.05). Post hoc analysis with a Bonferroni adjustment was used to determine between which segments these variations occurred.
\subsubsection{Questionnaires}
Participants completed three questionnaires. The first, an anonymous ethnographic questionnaire, was given in Stage 1. The second gauged interest in using avatars for therapy on a 5-point scale and was completed twice: after using pre-selected avatars in Stage 2 and after using self-created avatars in Stage 4. The third questionnaire in Stage 5 assessed immersion, amenity, availability, content quality, and overall satisfaction on a 5-point scale.
The second and third questionnaires were derived from a research paper that focuses on UX and experience questionnaires in Virtual Reality \cite{10.1145/2927929.2927955}.
\section{Findings: Thematic Analysis}
\subsection{Analysis of Interviews}
Key findings from thematic analysis are articulated in the following sub-sections. It should be noted that none of the participants had prior exposure to VR, for reasons that ranged from time limitations to lack of interest. This unique circumstance allowed their insights to unfold spontaneously, free from any preconceived notions or biases, which we believe adds a distinctive dimension to the research.
\newline \textbf{Avatar Creation:}
Out of the 10 participants, 5 opted to use the face of a renowned celebrity, while another 5 chose to alter the gender of the avatar they created. All participants expressed a desire to craft an avatar resembling their ideal self when it came to customization. Regarding their preference in therapy, every participant unanimously favored their customized avatar. This preference aligns with research indicating that personalized avatars significantly enhance body ownership, presence, and empowerment compared to generic avatars \cite{gall2021embodiment}. For 3 participants, this included wanting to express themselves through singing and dancing.
\begin{quote}
    \textit{"(translated from Japanese) I want to dance and sing when I'm like this (a custom avatar made with the face of their favorite singer) because I feel like I'm similar to my favorite singer now. I think this is cool for people who have a complex about their appearance and would be able to talk more freely to the doctor this way."}
\end{quote}
\textbf{Avatar Perception and Communication:}
When questioned about their choice of the pre-selected avatar, 8 participants cited its appeal due to cuteness, while the remaining 2 mentioned selecting it because it seemed enjoyable. 
However, when prompted to compare between the pre-selected and custom avatars, 8 out of 10 participants expressed a preference for the customized representations resembling their ideal selves. They indicated that such avatars enhance confidence and alleviate inhibitions, consistent with previous research findings \cite{messinger_relationship_2008}.
All participants expressed interest in enhancing the facial expressions of custom avatars, indicating a desire for greater emotional expressiveness. In contrast, when considering the pre-selected avatars, participants tended to prioritize enjoyment but perceived them as disconnected from their own selves.
\subsection{Average Questionnaire Results}
\begin{table}[]
\caption{Average 5 point score ratings for Questionnaire Answers} \label{tab:freq} \centering
\begin{tabular}{@{}llll@{}}
\toprule
\textbf{Therapy for VR} & \textbf{56} & \textbf{VR Experience} & \textbf{40}\\ \midrule
Good Idea & 4.15 & Amenity & 2.85 \\
Interest & 4.8 & Availability & 2.8\\
Enjoyment & 4.65 & Sense of Immersion & 3.5\\
Willingness to Use & 3.95 & Overall Satisfaction & 4.63\\ \bottomrule
\end{tabular}
\end{table}
\textbf{Avatars in Therapy:}
For both pre-selected and custom avatars, 7 out of 10 participants expressed positive views regarding the concept of VR therapy, considering it a good, fun, and intriguing idea they would be willing to engage with. Although there was a slightly stronger preference for custom avatars in the context of therapy, it's important to note that any disparities observed in the questionnaire responses may be influenced by the novelty of the VR experience, as all participants were first-time users.
\newline \textbf{VR Experience:}
Overall, participants rated the quality and satisfaction of the VR system as notably high, with a strong sense of immersion reported. However, the ease of use, in terms of amenity and availability, received an average rating. It's worth acknowledging that these ratings may be biased due to the fact that all 10 participants were experiencing VR for the first time.
\begin{figure}[htb]
    \centering  
    \includegraphics[width=1\linewidth]{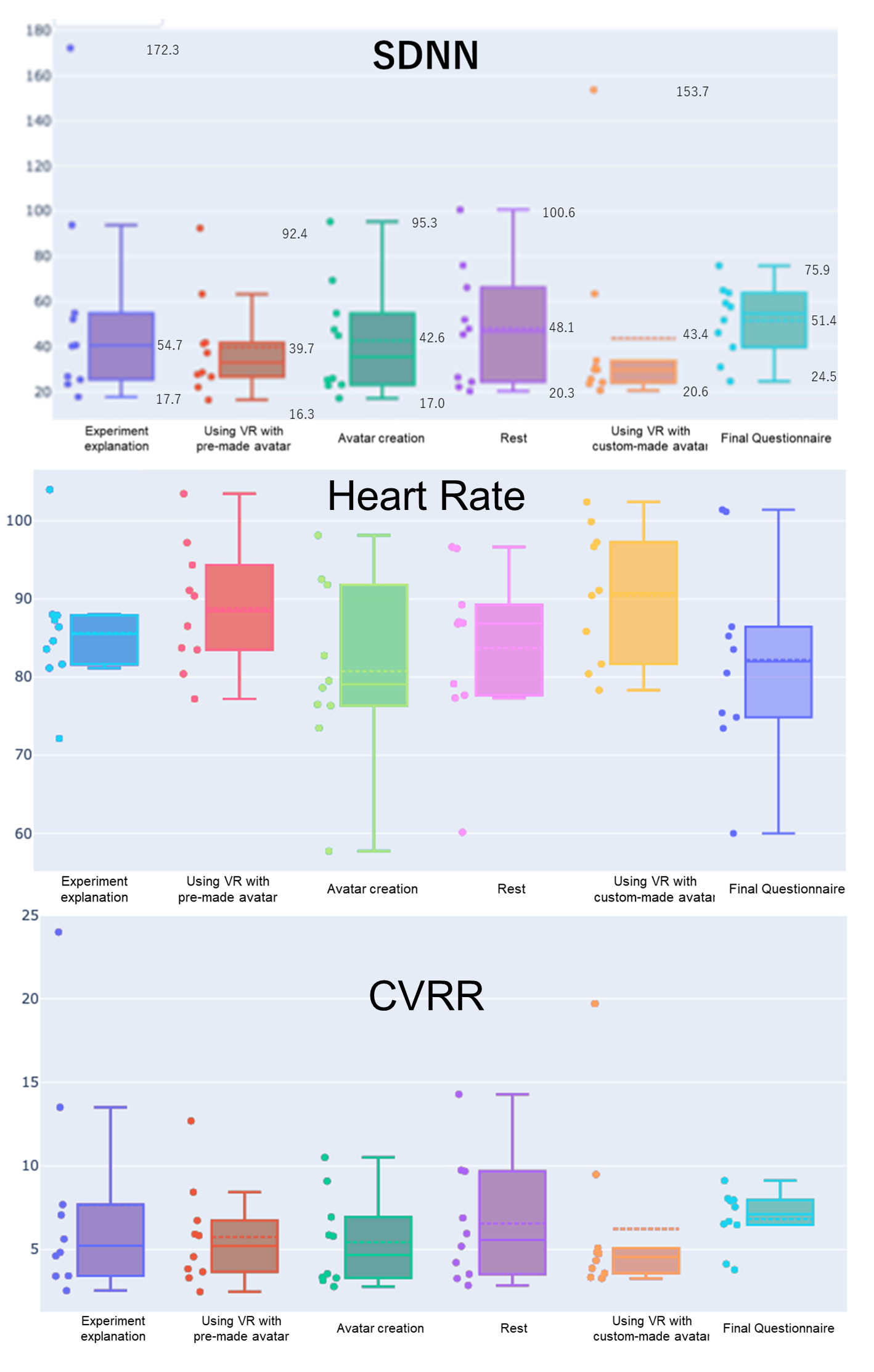} 
    \caption{Box plots of SDNN, Heart rate and CVRR of participants across the 6 activities featured in the experiment} 
    \label{fig:data}  
\end{figure}
\section{Findings: Physiological Data}
Results from the HRV analysis revealed consistent trends of greater physiological activation across participants during both VR sessions compared to the other periods, as indicated by increased heart rate and reduced SDNN and CVRR values (See Figure \ref{fig:data}). In particular, during the VR session featuring custom-made avatars participants displayed greater signs of arousal (Heart rate = 90.8, CVRR = 4.6, SDNN = 29.7), compared to the initial session in which they relied on pre-made avatars (Heart rate = 88.4, CVRR = 5.2, SDNN = 32.9). Significant difference were identified between the VR session with customised avatars and the rest period for Heart rate (p = .022), with the former associated with greater values. No other significant differences were found between resting periods and the VR session with customised avatars for the other two measures. However, CVRR differences in the values recorded for the VR session with pre-made vs customised avatars were close to significant level (p = .054). It is important to note that for all participants, the VR session with pre-made avatars was their first VR experience. This novelty likely increased their physiological responses due to excitement rather than a specific reaction to the avatars. This may explain the lack of significant differences between the two VR sessions, despite consistently higher arousal when using custom-made avatars.

\section{Conclusion}
The LGBTQ+ community grapples with social disparities and discrimination. Anonymity concerns can deter LGBTQ+ individuals from seeking therapy. However, the desire for representation is still crucial for members of the community, who struggle to express their identity in many situations. Virtual reality can support access to therapy in a way that preserves anonymity while offering an opportunity for self representation. Findings from the semi-structured interviews, questionnaires, and physiological signal analysis all point towards a preference for customized avatars in therapy. This consistent trend supports the argument for integrating an avatar creator before beginning VR therapy sessions. This approach allows individuals to tailor avatars to reflect their ideal selves, potentially enhancing their sense of confidence, immersion, and therapeutic engagement. Further studies should explore this opportunity in more depth, involving larger segments of the LGBTQ+ community.

\section{Acknowledgements}
Special thanks to Akta community center for their co-operation in the experiment. This work was supported by Council for Science, Technology and Innovation (CSTI), Cross-ministerial Strategic Innovation Promotion Program (SIP) and "Innovation of Inclusive Community Platform" (Grant Number JPJ012248, Funding Agency: National Institutes of Biomedical Innovation, Health and Nutrition (NIBIOHN)), JST Moonshot R\&D Program “Cybernetic being” Project (Grant number JPMJMS2013) and JST COI-NEXT “Minds1020Lab” Project (Grant Number JPMJPF2203).


\bibliographystyle{abbrv-doi}

\bibliography{template}
\end{document}